\documentclass[3p,times]{elsarticle}
\usepackage{amssymb}
\usepackage{epsfig}
\usepackage{mathrsfs}
\usepackage{graphicx}
\usepackage{amssymb}
\usepackage{amsmath}
\usepackage{lscape}
\usepackage{color}
\usepackage{dcolumn}
\usepackage{threeparttable}
\usepackage{booktabs}
\usepackage[labelfont=bf,singlelinecheck=false,font=footnotesize]{caption} 
\usepackage{hyperref}
\usepackage{multirow}

\biboptions{square,sort&compress} 
\definecolor{red}{rgb}{1,0,0}

\journal{Chemical Physics Letters}

\begin{document}
\begin{frontmatter}
\title{Extension of the CC($P$;$Q$) Formalism to the 
Electron Attachment and Ionization Potential
Equation-of-Motion Coupled-Cluster Frameworks}
\author[label1]{Karthik Gururangan\corref{cor1}}
\cortext[cor1]{Corresponding author}
\ead{gururang@msu.edu}
\author[label2]{Stephen H. Yuwono}
\ead{syuwono@fsu.edu}
\author[label2]{A. Eugene DePrince III}
\ead{adeprince@fsu.edu}
\author[label1,label3]{Piotr Piecuch}
\ead{piecuch@chemistry.msu.edu}
\address[label1]{Department of Chemistry, Michigan State University, 
East Lansing, Michigan 48824, USA}
\address[label2]{Department of Chemistry and Biochemistry, Florida State University, Tallahassee, Florida 32306, USA}
\address[label3]{Department of Physics and Astronomy, Michigan State University, East Lansing, Michigan 48824, USA}

\begin{abstract}
We combine the electron attachment (EA) and ionization potential (IP) 
equation-of-motion (EOM) coupled-cluster (CC) approaches with the 
CC($P$;$Q$) formalism. The resulting methodologies are used 
to describe the electronic states of several open-shell molecules, with 
the goal of approximating high-level EA/IP-EOMCC energetics 
corresponding to a full treatment of 3-particle--2-hole (3$p$-2$h$) and 
3-hole--2-particle (3$h$-2$p$) excitations on top of CC with singles and
doubles (CCSD). We show that the active-orbital-based EA/IP-EOMCC CC($P$;$Q$) 
approaches, abbreviated as \mbox{EA/IP-CC(t;3)}, achieve 
sub-millihartree
accuracies relative to the parent 
EA-EOMCCSD(3$p$-2$h$)/IP-EOMCCSD(3$h$-2$p$) data using reduced
computational effort, while improving upon their 
completely renormalized EA/IP-CR-EOMCC(2,3) counterparts.
\end{abstract}




\end{frontmatter}

\section{Introduction}
\label{sec1}
The single-reference coupled-cluster (CC) theory 
\cite{cizek1,cizek4}
and its equation-of-motion (EOM) \cite{eomcc3}
extension to electronically excited and electron attached and ionized states
are very useful tools for high-accuracy calculations in quantum chemistry.
In this Letter, we focus on the electron attachment (EA) and ionization potential (IP)
EOMCC approaches \cite{eaccsd1,eaccsd2,%
hirataeaip,%
ipccsd2,ipccsd3,ipccsd1,eaccsdt1,ipccsdt1,ipccsdt-3,%
gour1,gour2,gour3}, 
which provide effective and rigorously spin-adapted frameworks for describing
electronic states of open-shell species, such as
radicals, via electron addition to or electron removal from their
closed-shell counterparts, as well as directly accessing molecular
ionization and electron attachment spectra. 
In the EA/IP-EOMCC approaches,
the ground ($\mu = 0$) and excited ($\mu > 0$) states of the target ($N \pm 1$)-electron
species are given by
\begin{equation}
|\Psi_{\mu}^{(N \pm 1)}\rangle = R_{\mu}^{(\pm 1)} |\Psi_{0}^{(N)}\rangle,
\label{eq1}
\end{equation}
where 
\begin{equation}
|\Psi_{0}^{(N)}\rangle = e^{T}|\Phi\rangle
\label{eq2}
\end{equation}
is the CC ground state of the underlying $N$-electron system, with 
$T = \sum_{n=1}^{M_{T}} T_{n}$ representing the cluster operator
and $|\Phi\rangle$ designating the $N$-electron reference determinant, which serves as
a Fermi vacuum. The electron attaching
and ionizing operators, defined as 
$R_{\mu}^{(+1)} = \sum_{n=0}^{M_R} R_{\mu,(n+1)p\mbox{-}nh}$
and
$R_{\mu}^{(-1)} = \sum_{n=0}^{M_R} R_{\mu,(n+1)h\mbox{-}np}$,
respectively, generate the target $(N \pm 1)$-electron states 
via $(n+1)$-particle--$n$-hole [$(n+1)p$-$nh$] or $(n+1)$-hole--$n$-particle [$(n+1)h$-$np$]
excitations out of the
$N$-electron ground state $|\Psi_{0}^{(N)}\rangle$.
The many-body components of the
$T$, $R_{\mu}^{(+1)}$, and $R_{\mu}^{(-1)}$ operators are
\begin{equation}
T_{n} = \sum_{\substack{i_{1}<\ldots<i_{n} \\ a_{1}<\ldots<a_{n}}} t_{a_1\cdots a_n}^{i_1\cdots i_n} a^{a_{1}}\ldots a^{a_{n}}a_{i_{n}}\ldots a_{i_{1}},
\end{equation}
\begin{equation}
R_{\mu,(n+1)p\mbox{-}nh} = \sum_{\substack{j_{1}<\ldots<j_{n} \\ a<b_{1}<\ldots<b_{n}}} r_{a b_1\cdots b_n}^{\phantom{i}j_1\cdots j_n}(\mu)\: a^{a}a^{b_{1}}\ldots a^{b_{n}}a_{j_{n}}\ldots a_{j_{1}},
\end{equation}
and
\begin{equation}
R_{\mu,(n+1)h\mbox{-}np} = \sum_{\substack{i<j_{1}<\ldots<j_{n} \\ b_{1}<\ldots<b_{n}}} r_{\phantom{a}b_1\cdots b_n}^{ij_1\cdots j_n}(\mu)\: a^{b_{1}}\ldots a^{b_{n}}a_{j_{n}}\ldots a_{j_{1}}a_{i},
\end{equation}
respectively, where $a^{p}$ ($a_{p}$) is the fermionic creation (annihilation) operator
associated with spinorbital $|p\rangle$ and we adopt the usual notation in which
$i,j,k,\ldots$ ($a,b,c,\ldots$) denote the spinorbitals occupied (unoccupied) in $|\Phi\rangle$.
The parameters $M_T$ and $M_R$ that specify the truncations in $T$ and 
$R_{\mu}^{(\pm 1)}$, respectively, define the conventional hierarchies of EA/IP-EOMCC 
approximations. Setting $M_T = 2$ and $M_R = 1$, for example, 
describes the ($N \pm 1$)-electron
species using up to 2$p$-1$h$ or 2$h$-1$p$ excitations on top of the $N$-electron CC calculations
with singles and doubles (CCSD) \cite{ccsd,ccsd2}, resulting in the 
EA-EOMCCSD(2$p$-1$h$) \cite{eaccsd1,eaccsd2,hirataeaip} and 
IP-EOMCCSD(2$h$-1$p$) \cite{ipccsd1,ipccsd2,ipccsd3,hirataeaip} approaches, respectively. 
While EA-EOMCCSD(2$p$-1$h$) and IP-EOMCCSD(2$h$-1$p$) 
are computationally practical approximations, characterized
by costs scaling as $\mathscr{N}^6$
with the system size $\mathscr{N}$,
they can only describe open-shell states dominated by 
1$p$/1$h$ excitations. They fail to
provide meaningful results for states with
significant 2$p$-1$h$/2$h$-1$p$ contributions \cite{gour1,gour2,gour3,ehara,eaip_sac_pec,gold_ip,silver_ip}. 
Higher-level EA/IP-EOMCC 
approximations, such as EA-EOMCCSD(3$p$-2$h$) \cite{hirataeaip,gour1,gour2,gour3}
and IP-EOMCCSD(3$h$-2$p$) \cite{hirataeaip,gour1,gour2,gour3},
which incorporate the 3$p$-2$h$/3$h$-2$p$ excitations ($M_R = 2$) on top of CCSD ($M_T = 2$), or 
their EA-EOMCCSDT(3$p$-2$h$) \cite{hirataeaip,eaccsdt1} and 
IP-EOMCCSDT(3$h$-2$p$) \cite{hirataeaip,ipccsdt1}
extensions that replace the $N$-electron CCSD state with its more accurate
CCSDT \cite{ccfullt,ccfullt2} counterpart ($M_T = 3$),
are required to accurately describe radical excitation and 
electron attachment/ioniziation spectra, especially in the presence of significant 2$p$-1$h$/2$h$-1$p$ 
correlations, which often arise when examining satellite peaks
in photoelectron spectra \cite{gold_ip,silver_ip} and excited electronic states 
of radicals \cite{gour1,gour2,gour3,ehara,eaip_sac_pec}.
Unfortunately, higher-level EA/IP-EOMCC methods are not easily applicable to larger molecules
and basis sets due to the expensive computational costs, which scale as $\mathscr{N}^{7}$ for 
EA-EOMCCSD(3$p$-2$h$)/IP-EOMCCSD(3$h$-2$p$) and $\mathscr{N}^{8}$ for 
EA-EOMCCSDT(3$p$-2$h$)/IP-EOMCCSDT(3$h$-2$p$).

The present work aims to address this issue by developing new EA/IP-EOMCC
methodologies capable of accurately describing electronic states 
of radical species at the EA-EOMCCSD(3$p$-2$h$) and IP-EOMCCSD(3$h$-2$p$) 
levels at small fractions of the computational effort. Several approximate
treatments of 3$p$-2$h$/3$h$-2$p$ correlations
within the EA/IP-EOMCC frameworks have been explored in the past.
Among them are the active-space EA- and IP-EOMCCSDt methods 
\cite{gour1,gour2,gour3}, which aim at reducing the costs of EA-EOMCCSD(3$p$-2$h$) and IP-EOMCCSD(3$h$-2$p$) 
calculations by retaining only the subsets of dominant 3$p$-2$h$ and 3$h$-2$p$ 
excitations in the 
$R_{\mu,3p-2h}$ and $R_{\mu,3h-2p}$ components of $R_{\mu}^{(\pm 1)}$ selected with
the help of active orbitals. 
Alternatively, one may resort to iterative perturbative treatments \cite{ipccsdt-3}
or apply noniterative corrections to the low-level 
EA-EOMCCSD(2$p$-1$h$)/IP-EOMCCSD(2$h$-1$p$) energetics
due to 3$p$-2$h$ or 3$h$-2$p$ effects 
determined with the help of perturbation theory
\cite{stanton-gauss-ip,saeh_ip_1999,krylov-ip-correction}.
In this work, we choose yet another strategy
and extend the biorthogonal moment expansions
defining the CC($P$;$Q$) formalism
introduced in Refs.\ \cite{jspp-chemphys2012,jspp-jcp2012}
(cf.\ Refs.\ \cite{ren1,kkppeom2,crccl_jcp,crccl_molphys} for the prior work),
which have demonstrated considerable success in the context of particle-conserving CC/EOMCC calculations
\cite{jspp-chemphys2012,jspp-jcp2012,%
stochastic-ccpq-prl-2017,%
stochastic-ccpq-molphys-2020,%
cipsi-ccpq-2021,adaptiveccpq2023}, 
to the EA/IP-EOMCC regime. This allows us to formulate the noniterative corrections due to 
all or remaining 3$p$-2$h$ or 3$h$-2$p$ excitations neglected in the EA-EOMCCSD(2$p$-1$h$),
IP-EOMCCSD(2$h$-1$p$), EA-EOMCCSDt, and IP-EOMCCSDt calculations without having to rely on
traditional perturbation theory concepts. 
When applied to EA-EOMCCSD(2$p$-1$h$) and IP-EOMCCSD(2$h$-1$p$),
the CC($P$;$Q$) corrections developed in this study result in the
completely renormalized (CR) EA/IP-EOMCC methodologies
abbreviated as EA-CR-EOMCC(2,3) and IP-CR-EOMCC(2,3), respectively, 
which can be viewed as the EA/IP counterparts of the particle-conserving
CR-CC(2,3) \cite{crccl_jcp,crccl_molphys} and CR-EOMCC(2,3) \cite{crccl_molphys} 
approaches (cf.\ Ref.\ \cite{morris_crea} for the related work in the context
of nuclear physics). 
The proposed CC($P$;$Q$) corrections due to the 3$p$-2$h$ or 3$h$-2$p$ correlations
missing in the
active-space EA-EOMCCSDt and IP-EOMCCSDt calculations, abbreviated as
EA-CC(t;3) and IP-CC(t;3), respectively, are analogous to the 
CC(t;3) approach of Refs.\ \cite{jspp-chemphys2012,jspp-jcp2012}.
By examining the ground and excited electronic states of the
CH, CNC, C$_{2}$N, SH, N$_{3}$, and NCO radicals,
we demonstrate that the CC($P$;$Q$)-based 
EA/IP-CR-EOMCC(2,3) and EA/IP-CC(t;3) corrections
significantly improve the corresponding 
EA-EOMCCSD(2$p$-1$h$)/IP-EOMCCSD(2$h$-1$p$) and 
EA/IP-EOMCCSDt energetics when compared
to the parent EA-EOMCCSD(3$p$-2$h$)/IP-EOMCCSD(3$h$-2$p$) data.
In particular, the  
EA/IP-CC(t;3) approaches developed in this work are capable of
recovering the high-level EA-EOMCCSD(3$p$-2$h$)/IP-EOMCCSD(3$h$-2$p$)
energetics to sub-millihartree or even microhartree accuracy levels 
with considerably reduced computational effort.
\section{Theory}
\label{sec2}
In developing the EA/IP-CR-EOMCC(2,3) and EA/IP-CC(t;3) approaches,
we generalize the CC($P$;$Q$) moment expansions of 
Refs.\ \cite{jspp-chemphys2012,jspp-jcp2012} to the particle-nonconserving EA/IP-EOMCC regime.
In the existing particle-conserving CC($P$;$Q$) framework, we solve the CC/EOMCC equations 
in a subspace of the many-electron Hilbert space called the $P$ space and
correct the resulting energies for the correlation
effects not captured by the $P$-space CC/EOMCC calculations
with the help of another subspace of the Hilbert space
called the $Q$ space. In the EA/IP-EOMCC-based formulation of CC($P$;$Q$) 
proposed in this study, we proceed in a similar manner.
Thus, after deciding on the truncation in 
the cluster operator $T$ describing the $N$-electron reference system, we define 
the $P$ space $\mathscr{H}^{(P)}$
and $Q$ space $\mathscr{H}^{(Q)}$ as subspaces of the appropriate 
($N\pm 1)$-electron sector of the 
Fock space relevant to the electron attached/ionized wave functions
of interest.
The $P$ space
is spanned by the $(n+1)p$-$nh$ 
or $(n+1)h$-$np$ determinants that dominate the target
$(N\pm 1)$-electron states, whereas the companion $Q$ space consists of the
complementary $(n+1)p$-$nh$ or $(n+1)h$-$np$ determinants, which we use to correct the 
$P$-space EA/IP-EOMCC calculations for missing higher-order correlations.

In the EA/IP-EOMCC methods developed in this work, the underlying $N$-electron
reference system is described using CCSD, \emph{i.e.}, $T = T_{1} + T_{2}$, although
other truncations in $T$ could be considered as well.
After solving the CCSD equations for the $N$-electron ground
state, 
we diagonalize the 
similarity-transformed Hamiltonian 
$\overline{H} = e^{-T_{1}-T_{2}}He^{T_{1}+T_{2}}$ in the $(N\pm1)$-electron $P$ space, as defined above, to obtain the 
ground- and excited-state energies $E_{\mu}^{(N \pm 1)}(P)$ of the target electron attached/ionized
species and the associated 
right and left eigenvectors of $\overline{H}$ corresponding to the
\begin{equation}
R_{\mu}^{(\pm 1)}(P)=\sum_{|\Phi_{K}\rangle \in \mathscr{H}^{(P)}} r_{\mu,K} \,E_{K}^{(\pm 1)}
\label{Rop_p}
\end{equation}
and
\begin{equation}
L_{\mu}^{(\pm 1)}(P)=\sum_{|\Phi_{K}\rangle \in \mathscr{H}^{(P)}} l_{\mu,K} \,(E_{K}^{(\pm 1)})^{\dagger}
\label{Lop_p}
\end{equation}
operators,  
which define the ket states
$|\Psi_{\mu}^{(N \pm 1)}(P)\rangle = R_{\mu}^{(\pm 1)}(P) \,e^{T_{1} + T_{2}}|\Phi\rangle$
and their bra
$\langle \tilde{\Psi}_{\mu}^{(N \pm 1)}(P)| = \langle \Phi|L_{\mu}^{(\pm 1)}(P)\, e^{-T_{1} - T_{2}}$
counterparts satisfying the biorthonormality
condition 
$\langle \tilde{\Psi}_{\mu}^{(N \pm 1)}(P)| \Psi_{\nu}^{(N \pm 1)}(P)\rangle = 
\langle \Phi | L_{\mu}^{(\pm 1)}(P)\,R_{\nu}^{(\pm 1)}(P) | \Phi\rangle = \delta_{\mu,\nu}$
[the $E_{K}^{(\pm 1)}$s in Eqs.\
(\ref{Rop_p}) and (\ref{Lop_p}) are elementary $(n+1)p$-$nh$ or $(n+1)h$-$np$ 
excitation operators generating
$(N \pm 1)$-electron determinants  
$|\Phi_{K}\rangle = E_{K}^{(\pm 1)} |\Phi\rangle$]. 
The final CC($P$;$Q$) energies of the
$(N\pm1)$-electron species are calculated
as
\begin{equation}
E_{\mu}^{(N \pm 1)}(P;Q) = E_{\mu}^{(N \pm 1)}(P) + \delta_{\mu}^{(N \pm 1)}(P;Q),
\label{E_PQ}
\end{equation}
where the noniterative corrections $\delta_{\mu}^{(N \pm 1)}(P;Q)$ are defined as
\begin{equation}
\delta_{\mu}^{(N \pm 1)}(P;Q)=\sum_{|\Phi_K\rangle \in \mathscr{H}^{(Q)}} \ell_{\mu,K}^{(N \pm 1)}(P)\:\mathfrak{M}_{\mu,K}^{(N \pm 1)}(P),
\label{momex}
\end{equation}
with $\mathfrak{M}_{\mu,K}^{(N \pm 1)}(P) = \langle \Phi_{K} | \overline{H}\, R_{\mu}^{(\pm 1)}(P) | \Phi \rangle$ representing the EA/IP-EOMCC 
analogs of the CC/EOMCC moments 
considered in Refs.\ \cite{ren1,kkppeom2}. The coefficients $\ell_{\mu,K}^{(N \pm 1)}(P)$
multiplying moments $\mathfrak{M}_{\mu,K}^{(N\pm 1)}(P)$ in Eq.\ (\ref{momex})
are given by
$\ell_{\mu,K}^{(N \pm 1)}(P) = \langle \Phi | L_{\mu}^{(\pm 1)}(P) \,\overline{H} |\Phi_{K}\rangle / D_{\mu,K}^{(N \pm 1)}(P)$, with 
$D_{\mu,K}^{(N \pm 1)}(P) = E_{\mu}^{(N \pm 1)}(P) - \langle \Phi_{K} | \overline{H} | \Phi_{K}\rangle$
defining the corresponding Epstein--Nesbet (EN) energy denominators,
although one can also use their M{\o}ller--Plesset
(MP) analogs (cf.\ Refs.\ \cite{crccl_jcp,crccl_molphys,stochastic-ccpq-prl-2017}).

This study focuses on the CC($P$;$Q$) approaches aimed at accurately approximating the 
EA-EOMCCSD(3$p$-2$h$) and
IP-EOMCCSD(3$h$-2$p$) energetics at small fractions of the computational costs
by correcting the results of 
the EA/IP-EOMCC calculations 
with all 1$p$/1$h$, all
2$p$-1$h$/2$h$-1$p$, 
and, optionally, some 
3$p$-2$h$/3$h$-2$p$ excitations on top of CCSD
for the many-electron correlation effects associated with all
or the remaining 3$p$-2$h$/3$h$-2$p$ 
contributions using the moment expansions defined by Eq.\ (\ref{momex}).
In the case of the noniterative corrections 
$\delta_{\mu}^{(N + 1)}(P;Q)$ to EA-EOMCCSD(2$p$-1$h$) due to 3$p$-2$h$ correlations, which define the 
EA-CR-EOMCC(2,3) method, the $P$ space is spanned by all 1$p$ ($|\Phi^{a}\rangle$)
and 2$p$-1$h$ ($|\Phi^{ab}_{\phantom{i}j}\rangle$) determinants
and the corresponding $Q$ space consists of all 3$p$-2$h$ determinants 
($|\Phi_{\phantom{i}jk}^{abc}\rangle$). Similarly, the
IP-CR-EOMCC(2,3) approach, which corrects the IP-EOMCCSD(2$h$-1$p$) energetics
for 3$h$-2$p$ effects, uses a $P$ space spanned by all 1$h$ ($|\Phi_{i}\rangle$) and 2$h$-1$p$ ($|\Phi_{ij}^{\phantom{a}b}\rangle$) determinants and a $Q$ space consisting of all 3$h$-2$p$ determinants ($|\Phi_{ijk}^{\phantom{a}bc}\rangle$). In the case of the
active-orbital-based EA-CC(t;3) methodology, the $P$ space is defined as all $|\Phi^{a}\rangle$ and $|\Phi^{ab}_{\phantom{i}j}\rangle$ determinants and the subset of 
3$p$-2$h$ determinants selected according to the formula
$|\Phi_{\phantom{i}jk}^{\mathbf{A}bc}\rangle$, where the index $\mathbf{A}$ runs over active unoccupied spinorbitals defined by the user, and the $Q$ space consists of the remaining 3$p$-2$h$ determinants that are not included in the $P$ space. Finally, the IP-CC(t;3) approach adopts a $P$ space spanned by all $|\Phi_{i}\rangle$, all
$|\Phi_{ij}^{\phantom{a}b}\rangle$, and the subset of 3$h$-2$p$ determinants defined as
$|\Phi_{ij\mathbf{K}}^{\phantom{a}bc}\rangle$, where the index $\mathbf{K}$ denotes
the user-defined active occupied spinorbitals, and the $Q$ space is spanned by the remaining 3$h$-2$p$ determinants. 
It is worth mentioning that one could extend the EA/IP-based 
CC($P$;$Q$) methods
to capture the correlation effects
due to all or a subset of 4$p$-3$h$/4$h$-3$p$, 5$p$-4$h$/5$h$-4$p$, 
etc.\ excitations, in addition to all or some 
3$p$-2$h$/3$h$-2$p$ determinants,
provided that the truncation of the $N$-electron
cluster operator, $M_{T}$, is greater than or equal to the highest
value of $n$ characterizing the 
$(n+1)p$-$nh$/$(n+1)h$-$np$ determinants
included in $\mathscr{H}^{(P)}$ and 
$\mathscr{H}^{(Q)}$
(a condition required to retain size-intensivity \cite{gour1,gour2,gour3}).
For example, we could perform CC($P$;$Q$) calculations using an ($N \pm 1$)-electron
$P$ space spanned by all 1$p$/1$h$, 2$p$-1$h$/2$h$-1$p$, and a subset of 3$p$-2$h$/3$h$-2$p$ and 
4$p$-3$h$/4$h$-3$p$ excitations, and correct the results for the
missing 3$p$-2$h$/3$h$-2$p$ and 4$p$-3$h$/4$h$-3$p$ effects, 
if we used CCSDT ($M_{T}=3$) to describe the $N$-electron reference species.

In analogy to the CC($P$;$Q$) methods of 
Refs.\ \cite{jspp-chemphys2012,jspp-jcp2012,stochastic-ccpq-prl-2017,stochastic-ccpq-molphys-2020,cipsi-ccpq-2021,adaptiveccpq2023} and their CR-CC(2,3) and CR-EOMCC(2,3) predecessors, the EA/IP-CR-EOMCC(2,3) and EA/IP-CC(t;3) approaches offer considerable savings in the computational effort compared to their EA-EOMCCSD(3$p$-2$h$)/IP-EOMCCSD(3$h$-2$p$) parents. This is because the excitation spaces used in the iterative EA/IP-EOMCC steps contain none or only some 3$p$-2$h$/3$h$-2$p$ determinants and the noniterative corrections $\delta_{\mu}^{(N \pm 1)}(P;Q)$ require fewer CPU operations than a single iteration of EA-EOMCCSD(3$p$-2$h$)/IP-EOMCCSD(3$h$-2$p$) 
(cf.\ Table \ref{table1} for the computational costs 
characterizing the different EA/IP-EOMCC methods 
considered in this work).
In the next section, we assess how effective the EA/IP-CR-EOMCC(2,3) and EA/IP-CC(t;3) methods are in improving the underlying 
EA-EOMCCSD(2$p$-1$h$)/IP-EOMCCSD(2$h$-1$p$) and 
EA/IP-EOMCCSDt computations and recovering the parent
EA-EOMCCSD(3$p$-2$h$)/IP-EOMCCSD(3$h$-2$p$)
energetics. In doing so, we examine two variants of the
EA/IP-CR-EOMCC(2,3) and EA/IP-CC(t;3) approaches, termed 
EA/IP-CR-EOMCC(2,3)$_{\mathrm{X}}$ and EA/IP-CC(t;3)$_{\mathrm{X}}$, respectively,
where $\mathrm{X} = \mathrm{A}$ or $\mathrm{D}$. Following the naming convention adopted in Refs.\ \cite{crccl_jcp,crccl_molphys,stochastic-ccpq-prl-2017}, the methods labeled with subscript $\mathrm{D}$ use the EN-like energy denominators $D_{\mu,K}^{(N \pm 1)}(P)$ in Eq.\ (\ref{momex}) when evaluating corrections 
$\delta_{\mu}^{(N \pm 1)}(P;Q)$. The methods using subscript $\mathrm{A}$ replace the EN form of $D_{\mu,K}^{(N \pm 1)}(P)$ by its MP counterpart.
\section{Results and Discussion}
\label{sec3}
\subsection{Computational Details}
\label{sec3.1}
To assess the performance of the EA/IP-CR-EOMCC(2,3) and EA/IP-CC(t;3) approaches, we applied them, along with the underlying 
EA-EOMCCSD(2$p$-1$h$)/IP-EOMCCSD(2$h$-1$p$) and  EA/IP-EOMCCSDt methods and their EA-EOMCCSD(3$p$-2$h$)/IP-EOMCCSD(3$h$-2$p$) parents, to the ground and valence excited states of the CH, CNC, C$_{2}$N, SH, N$_{3}$, and NCO radical species. The results of our EA-EOMCC calculations for CH, CNC, and C$_{2}$N are reported in Tables \ref{table2} and \ref{table3}, whereas Tables \ref{table4} 
and \ref{table5} summarize the IP-EOMCC computations  for SH, N$_{3}$, and NCO. With 
the exception of SH, all of our EA/IP-EOMCC calculations correspond to adiabatic transitions; calculations for SH were performed at the ground-state equilibrium geometry. The nuclear geometries characterizing the ground and
excited states of CH and the ground-state S--H separation in SH were the same
as those used in Ref.\ \cite{gour3},
whereas the CNC, C$_{2}$N, N$_{3}$,
and NCO molecules were described using the linear geometries optimized with the EA  
(CNC and C$_{2}$N) or IP (N$_{3}$ and NCO) 
variants of the symmetry-adapted-cluster (SAC) configuration interaction (CI) approach \cite{sacci4} abbreviated as SDT-$R$ \cite{sacci_r} in Ref.\ \cite{ehara}. All EA-EOMCC calculations for CH, CNC, and C$_{2}$N reported in this work employed the restricted Hartree--Fock (RHF) orbitals of the underlying (CH)$^{+}$, (CNC)$^{+}$, and (C$_{2}$N)$^{+}$ cationic references, whereas the IP-EOMCC calculations for SH, N$_{3}$, and NCO used the RHF orbitals of the closed-shell (SH)$^{-}$, (N$_{3}$)$^{-}$, (NCO)$^{-}$ anions. 
Our calculations for CH employed the aug-cc-pVQZ basis set
\cite{ccpvnz,augccpvnz}, while
for SH, we used the aug-cc-pV(T+d)Z \cite{augccpvndz} basis for sulfur and aug-cc-pVTZ \cite{ccpvnz,augccpvnz} for hydrogen. Following Ref.\ \cite{ehara}, the CNC, C$_{2}$N, N$_{3}$, and NCO radicals were described using the DZP[4s2p1d] basis set \cite{dzp1,dzp2}. The active orbital spaces used to define the $R_{\mu,3p\mbox{-}2h}$ and $R_{\mu,3h\mbox{-}2p}$ components of 
$R_{\mu}^{(\pm 1)}$ entering the EA/IP-EOMCCSDt and subsequent EA/IP-CC(t;3) calculations were the same as those used in Refs.\ \cite{gour3,ehara}. Thus, the EA-EOMCCSDt and EA-CC(t;3) calculations for CH used an active space consisting of the lowest-energy $\pi$ orbitals unoccupied in (CH)$^{+}$, whereas the active spaces employed in the analogous calculations for CNC and C$_{2}$N were spanned by the four orbitals corresponding to the two lowest $\pi$ shells unoccupied in (CNC)$^{+}$ and (C$_{2}$N)$^{+}$. In the case of the IP-EOMCCSDt and IP-CC(t;3) computations for SH, N$_{3}$ and NCO, the active spaces consisted of
the highest-energy $\pi$ shell occupied in the respective parent (SH)$^{-}$, (N$_{3}$)$^{-}$, and (NCO)$^{-}$ anions. The lowest-energy orbitals corresponding to the 1s shells of carbon, oxygen, nitrogen, and sulfur atoms were frozen in all post-RHF steps.

All of the EA- and IP-EOMCC methods used in our calculations, including the new EA-CR-EOMCC(2,3), IP-CR-EOMCC(2,3), EA-CC(t;3), and IP-CC(t;3) approaches developed in this work and the EA-EOMCCSD(2$p$-1$h$), IP-EOMCCSD(2$h$-1$p$), EA-EOMCCSDt, IP-EOMCCSDt, EA-EOMCCSD(3$p$-2$h$),
and IP-EOMCCSD(3$h$-2$p$) methods, were implemented in the open-source CCpy package  available on GitHub \cite{CCpy-GitHub}. The required RHF orbitals and one- and two-electron molecular integrals were generated with GAMESS \cite{gamess}. GAMESS, especially its EA-EOMCCSD(2$p$-1$h$), IP-EOMCCSD(2$h$-1$p$), EA-EOMCCSDt, IP-EOMCCSDt, EA-EOMCCSD(3$p$-2$h$),
and IP-EOMCCSD(3$h$-2$p$) routines developed in Ref.\ \cite{gour3}, was also useful in testing our CCpy codes. Our EA- and IP-CR-EOMCC(2,3) routines in CCpy were also tested against their automated implementations obtained with \texttt{p$^\dagger$q} \cite{pdaggerq2}.

Our selection of the above six open-shell species is motivated by the previous studies reported in Refs.\ \cite{gour3,ehara},
which demonstrated that EA-EOMCCSD(3$p$-2$h$) and  IP-EOMCCSD(3$h$-2$p$) -- the parent approaches for all of the CC($P$;$Q$) schemes considered in this study -- can
provide electronic spectra that closely match the results obtained with other high-level methods, including multireference CI (CH and SH) and 
EA/IP SAC-CI SDT-$R$ (CNC, C$_{2}$N, N$_{3}$, and NCO). With the exception of the $B\:^{2}\Sigma^{-}$ state of C$_{2}$N, 
which is characterized by significant 3$p$-2$h$ correlations \cite{ehara}, the
EA-EOMCCSD(3$p$-2$h$)/IP-EOMCCSD(3$h$-2$p$) calculations also predict adiabatic excitation energies for the low-lying states of CH, CNC, C$_{2}$N, N$_{3}$, and NCO in good agreement with experiment, with errors ranging from 0.03 to 0.50 eV and an average deviation of 0.20 eV (cf.\ Refs.\ \cite{gour3,ehara}). These observations justify
EA-EOMCCSD(3$p$-2$h$) and IP-EOMCCSD(3$h$-2$p$) as benchmarks for evaluating the accuracy of their 
CC($P$;$Q$)-based approximations.
\subsection{Performance of the CC($P$;$Q$)-based EA/IP-EOMCC Methodologies}
\label{sec3.2}
We begin our discussion by examining the EA-EOMCC calculations reported in Tables \ref{table2} and \ref{table3}.
As previously shown in Refs.\ \cite{gour3,ehara}, the basic 
EA-EOMCCSD(2$p$-1$h$) method can only treat the 
ground states of CH, CNC, and C$_2$N with reasonable accuracy 
relative to EA-EOMCCSD(3$p$-2$h$) (to within 5 millihartree), but it
breaks down when describing the valence excited states, producing
errors on the order of 0.1 hartree, which lead to
adiabatic excitation energies that deviate from EA-EOMCCSD(3$p$-2$h$)
by $\sim$2–4 eV.
The failure of EA-EOMCCSD(2$p$-1$h$) is largely a consequence of neglecting the 3$p$-2$h$ effects, which are essential for accurately describing electronic states with significant 2$p$-1$h$ character. This can be mitigated by the noniterative EA-CR-EOMCC(2,3) corrections. As shown in Tables \ref{table2} and \ref{table3}, the $\sim$2--5 millihartree differences between the EA-EOMCCSD(2$p$-1$h$) and EA-EOMCCSD(3$p$-2$h$) energies of the ground states of CH, CNC, and C$_{2}$N are reduced by a factor of $\sim$2--3 when the EA-CR-EOMCC(2,3) corrections to EA-EOMCCSD(2$p$-1$h$) are employed. In the case of excited states, the improvements offered by the EA-CR-EOMCC(2,3) corrections are even more substantial. 
The massive errors of EA-EOMCCSD(2$p$-1$h$) relative to EA-EOMCCSD(3$p$-2$h$), which usually exceed 100 millihartree,
are significantly reduced to 17.332--31.850 millihartree for CH, 20.996--22.060 millihartree for CNC, and 20.003--29.695 millihartree for C$_{2}$N when using EA-CR-EOMCC(2,3)$_{\rm A}$.
The analogous errors characterizing the EA-CR-EOMCC(2,3)$_{\rm D}$ calculations, which are 0.115--12.505 millihartree for CH, 14.346--14.517 millihartree for CNC, and 18.601--33.535 millihartree for C$_{2}$N,
are similar or smaller.
The noniterative EA-CR-EOMCC(2,3)$_{\rm A}$ and EA-CR-EOMCC(2,3)$_{\rm D}$ corrections also improve the adiabatic excitation spectra resulting from the EA-EOMCCSD(2$p$-1$h$) calculations, reducing the $\sim$2--4 eV errors relative to EA-EOMCCSD(3$p$-2$h$) obtained with EA-EOMCCSD(2$p$-1$h$) to 0.45--0.84 and 0.03--0.97 eV, respectively. While the EA-CR-EOMCC(2,3) methodology helps temper the failures 
of EA-EOMCCSD(2$p$-1$h$),
the $\sim$20--30 millihartree errors relative to EA-EOMCCSD(3$p$-2$h$) characterizing the majority of our EA-CR-EOMCC(2,3) results remain. In analogy to the particle-conserving CC($P$;$Q$) framework 
\cite{jspp-chemphys2012,jspp-jcp2012,%
stochastic-ccpq-prl-2017,%
stochastic-ccpq-molphys-2020,%
cipsi-ccpq-2021,adaptiveccpq2023}, 
this can be attributed to the decoupling of the $R_{\mu,3p\mbox{-}2h}$ component of $R_{\mu}^{(+1)}$ from its lower-rank $R_{\mu,1p}$ and $R_{\mu,2p\mbox{-}1h}$ counterparts when the EA-CR-EOMCC(2,3) corrections are constructed. As shown in Tables \ref{table2} and \ref{table3}, this problem can be addressed by turning to the EA-CC(t;3) approach, which replaces 
EA-EOMCCSD(2$p$-1$h$) by EA-EOMCCSDt prior to determining the 3$p$-2$h$ corrections.

Upon examining the EA-EOMCCSDt and EA-CC(t;3) data reported in Tables \ref{table2} and \ref{table3}, the benefits offered by including the dominant 3$p$-2$h$ excitations selected via active orbitals in the iterative EA-EOMCC steps are readily apparent. 
The energies resulting from the EA-EOMCCSDt calculations for the ground and excited states of CH, CNC, and C$_{2}$N differ from their EA-EOMCCSD(3$p$-2$h$)
parents by 1.454--3.243, 1.414--1.910, and 0.543--1.606 millihartree, respectively, which in most cases is a massive improvement over EA-EOMCCSD(2$p$-1$h$). 
In fact, with the exception of the $X\:^{2}\Pi$ and $a\:^{4}\Sigma^{-}$ states of CH and the $X\:^{2}\Pi_{\rm g}$ state of CNC, EA-EOMCCSDt outperforms EA-CR-EOMCC(2,3) as well, which highlights the importance of incorporating the coupling between $R_{\mu,3p\mbox{-}2h}$ and its 
lower-rank $R_{\mu,1p}$ and $R_{\mu,2p\mbox{-}1h}$ counterparts when treating states characterized by stronger 2$p$-1$h$ correlations. The adiabatic excitation spectra of CH, CNC, and C$_{2}$N computed with EA-EOMCCSDt are also very accurate, with errors relative to full EA-EOMCCSD(3$p$-2$h$) not exceeding 0.05 eV (cf.\ Refs.\ \cite{gour3,ehara}). By correcting the EA-EOMCCSDt energetics for the remaining 3$p$-2$h$ correlations using EA-CC(t;3), we obtain a virtually perfect agreement with the parent EA-EOMCCSD(3$p$-2$h$) data. Indeed, when using the EA-CC(t;3)$_{\rm A}$ approach, the 
1--3 millihartree errors relative to EA-EOMCCSD(3$p$-2$h$) resulting from the EA-EOMCCSDt calculations for CH, CNC, and C$_{2}$N reduce to 0.221--0.412, 0.194--0.241, and 0.084--0.201 millihartree, respectively. 
The EA-CC(t;3)$_{\rm D}$ corrections to 
EA-EOMCCSDt are
even more accurate, reproducing the target EA-EOMCCSD(3$p$-2$h$) energetics to within tens of microhartree or less. Even the largest deviation of 0.106 millihartree, observed in the EA-CC(t;3)$_{\rm D}$ calculations for the $a\:^{4}\Sigma^{-}$ state of CH, is still an order of magnitude smaller than the error obtained with EA-EOMCCSDt.
It is important to emphasize that the EA-CC(t;3) calculations reported in Tables \ref{table2} and \ref{table3} are much less expensive than the parent EA-EOMCCSD(3$p$-2$h$) computations, while providing virtually identical energetics. For example, the $P$ spaces adopted in our EA-EOMCCSDt and EA-CC(t;3) computations for CNC and C$_{2}$N used only about 30\% of all $S_{z}=1/2$ 3$p$-2$h$ determinants. In the case of CH, the fractions of $S_{z}=1/2$ 3$p$-2$h$ determinants included in the corresponding $P$ spaces were even smaller (less than 5\% of all $S_{z}=1/2$ 3$p$-2$h$ determinants). As demonstrated in Ref.\ \cite{gour3}, the use of small fractions of 3$p$-2$h$ determinants in the iterative EA-EOMCCSDt steps translates into substantial savings in the computational costs relative to
the parent EA-EOMCCSD(3$p$-2$h$) approach. The EA-CC(t;3) corrections to EA-EOMCCSDt do not change this, since they involve fewer CPU operations than a single iteration of EA-EOMCCSD(3$p$-2$h$).

Similar accuracy patterns are observed in the results of the CC($P$;$Q$)-based IP-EOMCC calculations reported in Tables \ref{table4} and \ref{table5}. Much like EA-EOMCCSD(2$p$-1$h$), the basic IP-EOMCCSD(2$h$-1$p$) approach is generally incapable of providing an accurate description, especially when 2$h$-1$p$ effects become more substantial \cite{gour3,ehara}.
In the case of SH, the discrepancies between the IP-EOMCCSD(2$h$-1$p$) and parent IP-EOMCCSD(3$h$-2$p$) energetics are on the order of 0.1--0.2 hartree for all but the $X\:^{2}\Pi$ and $A\:^{2}\Sigma^{+}$ states, which are described more accurately by IP-EOMCCSD(2$h$-1$p$) since these two states are the only ones in Table \ref{table4} dominated by 1$h$ correlations. The same is true when examining the vertical excitation spectrum of SH obtained with IP-EOMCCSD(2$h$-1$p$), where the errors relative to IP-EOMCCSD(3$h$-2$p$) are 0.06 eV for the $A\:^{2}\Sigma^{+}$ state and 3.5--5.3 eV for the remaining excited states shown in Table \ref{table4} (cf.\ Ref.\ \cite{gour3}). In the case of the electronic states of N$_{3}$ and NCO reported in Table \ref{table5}, for which 2$h$-1$p$ effects are small, the IP-EOMCCSD(2$h$-1$p$) method 
fares better, providing adiabatic transition energies in good agreement (0.3 eV or less) with the IP-EOMCCSD(3$h$-2$p$) results \cite{ehara}, but this is partly due to a fortuitous cancellation of the 10--20 millihartree errors characterizing the IP-EOMCCSD(2$h$-1$p$) energies relative to IP-EOMCCSD(3$h$-2$p$). In analogy to the EA-EOMCC computations, we can improve the IP-EOMCCSD(2$h$-1$p$) results by applying the IP-CR-EOMCC(2,3) corrections due to 3$h$-2$p$ effects. 
Indeed, as shown in Table \ref{table4}, the errors characterizing
the IP-EOMCCSD(2$h$-1$p$) results for the $X\:^{2}\Pi$ 
and $A\:^{2}\Sigma^{+}$ states of SH
decrease by a factor of nearly three or more using either
variant of the IP-CR-EOMCC(2,3) approach.
The error reductions offered by the IP-CR-EOMCC(2,3) corrections are even larger when 
examining
the remaining excited states of SH considered in Table \ref{table4}, 
which are
characterized by significant 2$h$-1$p$ contributions. 
For these states, the
137.280--203.145 millihartree
errors relative to IP-EOMCCSD(3$h$-2$p$) obtained with
IP-EOMCCSD(2$h$-1$p$) reduce
in the IP-CR-EOMCC(2,3)$_{\rm A}$ and IP-CR-EOMCC(2,3)$_{\rm D}$ calculations to 0.359--39.214 and 3.201--64.470 millihartree, respectively. 
As a result, the vertical excitation energies of SH obtained 
with IP-CR-EOMCC(2,3) greatly improve upon IP-EOMCCSD(2$h$-1$p$), agreeing with their IP-EOMCCSD(3$h$-2$p$) parent values
to within 0.04--1.71 eV.
For N$_{3}$ and NCO, the improvements
offered by the 3$h$-2$p$ corrections to IP-EOMCCSD(2$h$-1$p$) are 
less dramatic, since the electronic states of these two molecules 
included in Table \ref{table5} are dominated by 1$h$ excitations 
out of (N$_{3}$)$^{-}$ and (NCO)$^{-}$, but the reductions in 
errors relative to IP-EOMCCSD(3$h$-2$p$), from 9.587--20.154 
millihartree obtained with IP-EOMCCSD(2$h$-1$p$) to 0.808--7.044 
millihartree obtained with variants A and D of IP-CR-EOMCC(2,3), 
are still quite substantial. 
We can conclude that the 
IP-CR-EOMCC(2,3) corrections are helpful, but, as shown in Tables 
\ref{table4} and \ref{table5}, they are not as effective in 
approximating the IP-EOMCCSD(3$h$-2$p$) energetics as the
IP-CC(t;3) corrections to IP-EOMCCSDt.

The IP-CC(t;3) results reported in Tables \ref{table4} and \ref{table5} and their IP-EOMCCSDt counterparts clearly show the benefits of including the leading 3$h$-2$p$ determinants in the relevant $P$ spaces prior to determining the 3$h$-2$p$ corrections. Indeed, with the exception of the $A\:^{2}\Sigma^{+}$ state of NCO, the IP-EOMCCSDt approach substantially improves the IP-EOMCCSD(2$h$-1$p$) energetics, reducing the 9.898--203.145, 13.078--14.623, and 9.587--20.154 millihartree errors relative to IP-EOMCCSD(3$h$-2$p$) obtained for SH, N$_{3}$, and NCO with IP-EOMCCSD(2$h$-1$p$) to 0.015--1.003, 1.400--6.215, and 1.926--2.205 millihartree, respectively. In the case of the $A\:^{2}\Sigma^{+}$ state of NCO, the improvement over the basic IP-EOMCCSD(2$h$-1$p$) level offered by IP-EOMCCSDt is minimal, but this is no longer an issue when the IP-CC(t;3)$_{\rm A}$ and IP-CC(t;3)$_{\rm D}$ corrections to IP-EOMCCSDt are employed. The results in Tables \ref{table4} and \ref{table5} clearly show that the IP-CC(t;3) corrections not only improve the IP-EOMCCSDt energetics, but they are also considerably more accurate than their IP-EOMCCSD(2$h$-1$p$)-based IP-CR-EOMCC(2,3) counterparts, and this applies to all of the electronic states of SH, N$_{3}$, and NCO considered in this work. 
For SH, both IP-CC(t;3)$_{\rm A}$ and IP-CC(t;3)$_{\rm D}$ recover the parent IP-EOMCCSD(3$h$-2$p$) data to within small fractions of a millihartree. In the cases of N$_{3}$ and NCO, the IP-CC(t;3)$_{\rm A}$ method shows a maximum error relative to IP-EOMCCSD(3$h$-2$p$) of just 1.690 millihartree, while the more complete IP-CC(t;3)$_{\rm D}$ approach
recovers its parent energetics to within one millihartree for 
all electronic states.
In analogy to the EA-CC(t;3) approach, the IP-CC(t;3) calculations
provide IP-EOMCCSD(3$h$-2$p$)-level results with reduced
computational costs.
Indeed, the $P$ spaces entering our IP-EOMCCSDt and IP-CC(t;3) 
calculations for SH, N$_{3}$, and NCO 
include approximately 
60\% of all $S_{z} = 1/2$
3$h$-2$p$ determinants, which
translate to sizeable speedups compared to the full 
IP-EOMCCSD(3$h$-2$p$)
computations. 
In fact, 
the iterative IP-EOMCCSDt and noniterative IP-CC(t;3) steps require CPU efforts that are comparable to the underlying $N$-electron CCSD calculations.
\section{Summary}
\label{sec4}
In this work, we have 
developed the CC($P$;$Q$)-based
EA/IP-EOMCC approaches that correct the energetics resulting from
the EA-EOMCCSD(2$p$-1$h$) and
IP-EOMCCSD(2$h$-1$p$) calculations as well as their 
active-space EA/IP-EOMCCSDt
extensions for all or the remaining 3$p$-2$h$ or 3$h$-2$p$ correlation
effects. While the EA/IP-CR-EOMCC(2,3) methods provide
sizeable improvements over the low-level EA-EOMCCSD(2$p$-1$h$)
and IP-EOMCCSD(2$h$-1$p$) data, the most accurate results 
are obtained using active-orbital-based EA/IP-CC(t;3) calculations,
which are capable of recovering the parent 
EA-EOMCCSD(3$p$-2$h$)/IP-EOMCCSD(3$h$-2$p$) energetics to within 
a fraction of a millihartree using reduced computational effort.
In the future, it would be interesting to extend the present
CC($P$;$Q$) formalism to target 
higher-level EA/IP-EOMCC approaches,
such as EA-EOMCCSDT(3$p$-2$h$) and IP-EOMCCSDT(3$h$-2$p$),
in a computationally efficient manner.
This may be accomplished,
for example, by replacing the $N$-electron 
CCSD similarity-transformed Hamiltonian with its more accurate 
CCSDt \cite{semi0b,semi2,semi4} counterpart.
Furthermore, we can combine 
the 
EA/IP-EOMCC-based CC($P$;$Q$) framework with the 
semi-stochastic \cite{stochastic-ccpq-prl-2017}, selected-CI-based \cite{cipsi-ccpq-2021}, and 
adaptive CC($P$;$Q$) \cite{adaptiveccpq2023}
techniques, to automate the construction of the 
$P$ and $Q$ spaces entering the CC($P$;$Q$) computations.
\section*{CRediT authorship contribution statement}
\textbf{Karthik Gururangan}: Methodology, Software, Data curation, Formal analysis, Validation, Writing - original
draft. \textbf{Stephen H. Yuwono}: Methodology, Software, Data curation, Formal analysis, Validation, Writing - original draft. 
\textbf{A. Eugene DePrince III}: Funding acquisition, Project administration, Resources, Supervision, Writing - reviewing and editing.
\textbf{Piotr
Piecuch}: Conceptualization, Methodology, Formal analysis, Investigation, Funding acquisition, Project administration, Resources, Supervision, Validation, Writing - reviewing and editing.
\section*{Declaration of competing interest}
The authors declare that they have no known competing financial
interests or personal relationships that could have appeared to influence
the work reported in this paper.
\section*{Data availability}
The data that support the findings of this study are available
within the article.
\section*{Acknowledgments}
This work has been supported by the Chemical Sciences,
Geosciences and Biosciences Division, 
Office of Basic Energy Sciences, 
Office of Science, 
U.S. Department of Energy 
(Grant No. DE-FG02-01ER15228 to P.P). A.E.D. acknowledges support by the Office of Advanced Scientific Computing Research and Office of Basic Energy Sciences, Scientific Discovery through the Advanced Computing (SciDAC) program, Office of Science, U.S. Department of Energy, under Award No. DE-SC0022263.
%

%
\newpage
\clearpage

\begin{table*}[h]
\centering
\footnotesize
\begin{threeparttable}
\caption{Summary of the
EA/IP-EOMCC methodologies employed in this work
using the language of the CC($P$;$Q$) formalism, including the 
$P$ spaces $\mathscr{H}^{(P)}$ adopted in the iterative EA/IP-EOMCC calculations
and the $Q$ spaces $\mathscr{H}^{(Q)}$ defining the relevant noniterative
$\delta_{\mu}^{(N\pm 1)}(P;Q)$ corrections (if any), along with the corresponding 
computational costs.
In all cases, the reported computational costs reflect the number
of operations needed to evaluate the most expensive terms
entering the iterative and (if applicable) noniterative steps of 
each EA/IP-EOMCC-based CC($P$;$Q$) calculation.
Computational costs are expressed in terms of $n_\mathrm{o}$ 
and $n_\mathrm{u}$, which
respectively represent the numbers of correlated spinorbitals that are 
occupied or unoccupied in the underlying $N$-electron reference
$|\Phi\rangle$.
In reporting the costs of the active-orbital-based EA-EOMCCSDt
and EA-CC(t;3) approaches, we denote the number of spinorbitals
unoccupied in $|\Phi\rangle$ chosen as active by 
$N_\mathrm{u}$ ($1 \le N_\mathrm{u} \le n_\mathrm{u}$). Similarly, the number of occupied spinorbitals
that are active in the IP-EOMCCSDt and IP-CC(t;3) computations
is given by $N_\mathrm{o}$ ($1 \le N_\mathrm{o} \le n_\mathrm{o}$).}
\label{table1}
\begin{tabular}{lcccc}
\hline\hline
\multirow{2}{*}{Method} &
\multirow{2}{*}{$\mathscr{H}^{(P)}$ space\tnote{a}} &
\multirow{2}{*}{$\mathscr{H}^{(Q)}$ space\tnote{b}} &
\multicolumn{2}{c}{Computational Cost} \\
\cline{4-5}
 & & & Iterative\tnote{c} & Noniterative\tnote{d} \\
\hline
EA-EOMCCSD(2$p$-1$h$) & $\mathrm{span}\{|\Phi^{a}\rangle,|\Phi_{\phantom{i}j}^{ab}\rangle\}$ & $\{\varnothing\}$ & $n_\mathrm{o} n_\mathrm{u}^4$ \tnote{e} & --- \\
EA-CR-EOMCC(2,3)      & $\mathrm{span}\{|\Phi^{a}\rangle,|\Phi_{\phantom{i}j}^{ab}\rangle\}$ & $\mathrm{span}\{|\Phi_{\phantom{i}jk}^{abc}\rangle\}$ & $n_\mathrm{o} n_\mathrm{u}^4$ \tnote{e} & $n_\mathrm{o}^2 n_\mathrm{u}^4$ \\
EA-EOMCCSDt           & $\mathrm{span}\{|\Phi^{a}\rangle,|\Phi_{\phantom{i}j}^{ab}\rangle,|\Phi_{\phantom{i}jk}^{\mathbf{A}bc}\rangle\}$ & $\{\varnothing\}$ & $N_\mathrm{u} n_\mathrm{o}^2 n_\mathrm{u}^4$ & --- \\
EA-CC(t;3)            & $\mathrm{span}\{|\Phi^{a}\rangle,|\Phi_{\phantom{i}j}^{ab}\rangle,|\Phi_{\phantom{i}jk}^{\mathbf{A}bc}\rangle\}$ & $\mathrm{span}\{|\Phi_{\phantom{i}jk}^{abc}\rangle\} \ominus \mathrm{span}\{|\Phi_{\phantom{i}jk}^{\mathbf{A}bc}\rangle\}$ & $N_\mathrm{u} n_\mathrm{o}^2 n_\mathrm{u}^4$ & $(n_\mathrm{u} - N_\mathrm{u})^3 N_\mathrm{u} n_\mathrm{o}^2 n_\mathrm{u}$ \\
EA-EOMCCSD(3$p$-2$h$) & $\mathrm{span}\{|\Phi^{a}\rangle,|\Phi_{\phantom{i}j}^{ab}\rangle,|\Phi_{\phantom{i}jk}^{abc}\rangle\}$ & $\{\varnothing\}$ & $n_\mathrm{o}^2 n_\mathrm{u}^5 $ & --- \\
\\
IP-EOMCCSD(2$h$-1$p$) & $\mathrm{span}\{|\Phi_{i}\rangle,|\Phi_{ij}^{\phantom{a}b}\rangle\}$ & $\{\varnothing\}$ & $n_\mathrm{o}^3 n_\mathrm{u}^2$ \tnote{e} & --- \\
IP-CR-EOMCC(2,3)      & $\mathrm{span}\{|\Phi_{i}\rangle,|\Phi_{ij}^{\phantom{a}b}\rangle\}$ & $\mathrm{span}\{|\Phi_{ijk}^{\phantom{a}bc}\rangle\}$ & $n_\mathrm{o}^3 n_\mathrm{u}^2$ \tnote{e} & $n_\mathrm{o}^3 n_\mathrm{u}^3$ \\
IP-EOMCCSDt           & $\mathrm{span}\{|\Phi_{i}\rangle,|\Phi_{ij}^{\phantom{a}b}\rangle,|\Phi_{ij\mathbf{K}}^{\phantom{a}bc}\rangle\}$ & $\{\varnothing\}$ & $N_\mathrm{o} n_\mathrm{o}^2 n_\mathrm{u}^4$ & --- \\
IP-CC(t;3)            & $\mathrm{span}\{|\Phi_{i}\rangle,|\Phi_{ij}^{\phantom{a}b}\rangle,|\Phi_{ij\mathbf{K}}^{\phantom{a}bc}\rangle\}$ & $\mathrm{span}\{|\Phi_{ijk}^{\phantom{a}bc}\rangle\} \ominus \mathrm{span}\{|\Phi_{ij\mathbf{K}}^{\phantom{a}bc}\rangle\}$ & $N_\mathrm{o} n_\mathrm{o}^2 n_\mathrm{u}^4$ & $(n_\mathrm{o} - N_\mathrm{o})^3 N_\mathrm{o} n_\mathrm{u}^3$ \\
IP-EOMCCSD(3$h$-2$p$) & $\mathrm{span}\{|\Phi_{i}\rangle,|\Phi_{ij}^{\phantom{a}b}\rangle,|\Phi_{ijk}^{\phantom{a}bc}\rangle\}$ & $\{\varnothing\}$ & $n_\mathrm{o}^3 n_\mathrm{u}^4 $ & --- \\
\hline\hline
\end{tabular}
\begin{tablenotes}
\footnotesize
\item[a]{
The ($N \pm 1)$-electron $P$ space used to define 
the iterative steps of the CC($P$;$Q$)
calculation.
}
\item[b]{
The ($N \pm 1)$-electron $Q$ space used to define 
the noniterative steps of the CC($P$;$Q$)
calculation.
}
\item[c]{
The most expensive steps in solving the iterative EA/IP-EOMCC equations in the
$P$ space.
}
\item[d]{
The most expensive steps involved in computing the $\delta_{\mu}^{(N\pm 1)}(P;Q)$ corrections to the energies
resulting from the $P$-space
EA/IP-EOMCC calculations.
}
\item[e]{
The computational cost associated with solving the $P$-space EA/IP-EOMCC
equations is less expensive compared to the underlying $N$-electron 
CCSD calculation, which requires  
$n_\mathrm{o}^2 n_\mathrm{u}^4$ CPU operations.
}
\end{tablenotes}
\end{threeparttable}
\end{table*}
%
%
\begin{table*}[h]
\centering
\footnotesize
\begin{threeparttable}
\caption{
The total electronic energies obtained using the EA-EOMCCSD(2$p$-1$h$), EA-CR-EOMCC(2,3),
EA-EOMCCSDt, EA-CC(t;3), and EA-EOMCCSD(3$p$-2$h$) approaches for the ground and
low-lying excited states of
the ${\rm CH}$ radical, as described using the aug-cc-pVQZ \cite{ccpvnz,augccpvnz} basis set. 
The EA-EOMCCSD(3$p$-2$h$) results are given in hartree and the EA-EOMCCSD(2$p$-1$h$),
EA-CR-EOMCC(2,3), EA-EOMCCSDt, and EA-CC(t;3) data are reported as errors relative
to EA-EOMCCSD(3$p$-2$h$) in millihartree. 
The EA-EOMCC calculations for the ground and excited states of CH employed
the RHF orbitals of the corresponding closed-shell $\mathrm{(CH)}^{+}$ cation.
In all post-RHF steps, the lowest-energy core orbital correlating
with the 1s shell of carbon was kept frozen.
}
\label{table2}
\begin{tabular}{lccccccc}
\hline\hline
& & \multicolumn{2}{c}{EA-CR-EOMCC(2,3)$_{\rm X}$} & & \multicolumn{2}{c}{EA-CC(t;3)$_{\rm X}$\tnote{b}} & \\
\cline{3-4} \cline{6-7}
State\tnote{a} & EA-EOMCCSD(2$p$-1$h$) & $\rm{X} = \rm{A}$ & $\rm{X} = \rm{D}$ & EA-EOMCCSDt\tnote{b} & $\rm{X} = \rm{A}$ & \rm{X} = \rm{D} & EA-EOMCCSD($3p\mbox{-}2h$) \\
\hline
${X\:^{2}}\Pi$ & 2.173 & 0.879 & 1.070 & 3.243 & 0.412 & 0.076 & $-$38.418 732 \\
${a\:^{4}}\Sigma^{-}$ & 74.764 & 22.057 & 0.115 & 1.454 & 0.251 & 0.106 & $-$38.390 598 \\
${A\:^{2}}\Delta$ & 84.101 & 20.292 & $-$8.279 & 1.720 & 0.221 & 0.039 & $-$38.310 249 \\
${B\:^{2}}\Sigma^{-}$ & 117.876 & 31.850 & $-$12.505 & 1.683 & 0.239 & 0.063 & $-$38.294 158 \\
${C\:^{2}}\Sigma^{+}$ & 69.816 & 17.332 & $-$4.004 & 2.565 & 0.265 & $-$0.010 & $-$38.270 359\\
\hline\hline
\end{tabular}
\begin{tablenotes}
\footnotesize
\item[a]{
For each electronic state of CH, the EA-EOMCC calculations were performed 
at the corresponding equilibrium C--H separation, which is the same as that used in Ref.\ \cite{gour3}.
}
\item[b]{
The active space defining the EA-EOMCCSDt and 
EA-CC(t;3)$_{\mathrm{X}}$ calculations 
consisted of the lowest-energy $\pi$ orbitals unoccupied in (CH)$^{+}$.
}
\end{tablenotes}
\end{threeparttable}
\end{table*}
%
\begin{table*}[h]
\centering
\footnotesize
\begin{threeparttable}
\caption{
The total electronic energies obtained using the EA-EOMCCSD(2$p$-1$h$), EA-CR-EOMCC(2,3),
EA-EOMCCSDt, EA-CC(t;3), and EA-EOMCCSD(3$p$-2$h$) approaches for the ground
and low-lying excited states of the CNC and $\rm{C_2}$N molecules,
as described using the DZP[4s2p1d] basis set of Refs.\ \cite{dzp1,dzp2}.
The EA-EOMCCSD(3$p$-2$h$) results are given in hartree and the EA-EOMCCSD(2$p$-1$h$),
EA-CR-EOMCC(2,3), EA-EOMCCSDt, and EA-CC(t;3) data are reported as errors relative
to EA-EOMCCSD(3$p$-2$h$) in millihartree.
The RHF orbitals of the closed-shell $\rm{(CNC)}^{+}$ and $\rm{(C_2N)}^{+}$ cation
references were used in the EA-EOMCC calculations for CNC and $\rm{C_2}$N,
respectively. The lowest-energy orbitals that correlate with the 
1s shell of carbon and nitrogen were kept frozen. 
}
\label{table3}
\begin{tabular}{llccccccc}
\hline\hline
& & & \multicolumn{2}{c}{EA-CR-EOMCC(2,3)$_{\rm X}$} & & \multicolumn{2}{c}{EA-CC(t;3)$_{\rm X}$\tnote{b}} & \\
\cline{4-5} \cline{7-8}
Molecule\tnote{a} & State & EA-EOMCCSD(2$p$-1$h$) & $\rm{X} = \rm{A}$ & $\rm{X} = \rm{D}$ & EA-EOMCCSDt\tnote{b} & $\rm{X} = \rm{A}$ & \rm{X} = \rm{D} & EA-EOMCCSD($3p\mbox{-}2h$) \\
\hline
%
CNC & ${X\:^{2}}\Pi_{\rm g}$        & 4.881   & 2.057  & 1.658     & 1.910 & 0.241 & $-$0.022 & $-$130.411 530\\
    & ${A\:^{2}}\Delta_{\rm u}$     & 118.827 & 22.060 & $-$14.517 & 1.177 & 0.194 & 0.002    & $-$130.260 674\\
    & ${B\:^{2}}\Sigma_{\rm u}^{+}$ & 112.241 & 20.996 & $-$14.346 & 1.414 & 0.208 & $-$0.022 & $-$130.238 150 
\vspace{1em} \\
$\rm{C_2}$N & ${X\:^{2}}\Pi$ & 4.696   & 2.708  & 2.019     & 1.606 & 0.201 & $-$0.020 & $-$130.404 919\\
    & ${A\:^{2}}\Delta$      & 119.913 & 21.771 & $-$18.601 & 0.612 & 0.092 & 0.008    & $-$130.292 647\\
    & ${B\:^{2}}\Sigma^{-}$  & 158.282 & 29.695 & $-$33.535 & 0.543 & 0.084 & 0.009    & $-$130.269 785\\
    & ${C\:^{2}}\Sigma^{+}$  & 111.733 & 20.003 & $-$19.099 & 0.824 & 0.107 & $-$0.008 & $-$130.264 924\\
\hline\hline
\end{tabular}
\begin{tablenotes}
\footnotesize
\item[a]{
Both the CNC and $\rm{C_2}$N molecules employed linear nuclear geometries, with
C--N and C--C bond lengths taken from Ref.\ \cite{ehara}, in which
they were optimized for each electronic state using 
the EA SAC-CI SDT-$R$ method.
}
\item[b]{
The active spaces used in the EA-EOMCCSDt and EA-CC(t;3)$_{\mathrm{X}}$ calculations
for CNC and $\rm{C_2}$N were spanned by the four orbitals corresponding to the two lowest $\pi$ shells unoccupied in (CNC)$^{+}$ and (C$_{2}$N)$^{+}$, respectively.
}
\end{tablenotes}
\end{threeparttable}
\end{table*}
\begin{table*}[h]
\centering
\footnotesize
\begin{threeparttable}
\caption{
The total electronic energies obtained using the IP-EOMCCSD(2$h$-1$p$), IP-CR-EOMCC(2,3),
IP-EOMCCSDt, IP-CC(t;3), and IP-EOMCCSD(3$h$-2$p$) approaches for the ground and
selected excited states of
the SH radical, as described using the aug-cc-pV(T+d)Z basis 
set \cite{augccpvndz} for S
and the aug-cc-pVTZ basis \cite{ccpvnz,augccpvnz} for H.
The IP-EOMCCSD(3$h$-2$p$) results are given in hartree and the IP-EOMCCSD(2$h$-1$p$),
IP-CR-EOMCC(2,3), IP-EOMCCSDt, and IP-CC(t;3) data are reported as errors relative
to IP-EOMCCSD(3$h$-2$p$) in millihartree.
The IP-EOMCC calculations were performed 
at the equilibrium S--H separation characterizing the ground state of SH, which is the same as 
that used in Ref.\ \cite{gour3},
and the RHF orbitals of the underlying closed-shell $\mathrm{(SH)}^{-}$ anion were employed throughout.
In all post-RHF steps, the lowest-energy core orbital correlating
with the 1s shell of sulfur was kept frozen.
}
\label{table4}
\begin{tabular}{lccccccc}
\hline\hline
& & \multicolumn{2}{c}{IP-CR-EOMCC(2,3)$_{\rm X}$} & & \multicolumn{2}{c}{IP-CC(t;3)$_{\rm X}$\tnote{a}} & \\
\cline{3-4} \cline{6-7}
State & IP-EOMCCSD(2$h$-1$p$) & $\rm{X} = \rm{A}$ & $\rm{X} = \rm{D}$ & IP-EOMCCSDt\tnote{a} & $\rm{X} = \rm{A}$ & \rm{X} = \rm{D} & IP-EOMCCSD($3h\mbox{-}2p$) \\
\hline
${X\:^{2}}\Pi$        & 9.898 & 2.904     & 1.576     & 0.309    & 0.034 & 0.031 & $-$398.388 242 \\
${A\:^{2}}\Sigma^{+}$ & 12.149 & 4.449     & 3.201     & $-$0.217 & 0.134 & 0.095    & $-$398.243 962 \\
${1\:^{4}}\Sigma^{-}$ & 203.145  & 19.235  & 64.350 & 0.030    & 0.006 & 0.007    & $-$398.167 978 \\
${1\:^{2}}\Sigma^{-}$ & 161.002  & $-$20.923 & 10.835 & 0.015    & 0.002 & 0.002    & $-$398.139 560 \\
${1\:^{2}}\Delta$     & 199.497  & 0.359 & 56.308 & 0.049    & 0.014 & 0.013    & $-$398.116 100 \\
${B\:^{2}}\Sigma^{+}$ & 154.422  & 39.214    & 64.470    & 1.003    & 0.149 & 0.151    & $-$398.084 521 \\
${1\:^{4}}\Pi$        & 137.280  & $-$28.023    & 6.658 & 0.057   & 0.016 & 0.016    & $-$398.073 621 \\
\hline\hline
\end{tabular}
\begin{tablenotes}
\footnotesize
\item[a]{
The active space defining the IP-EOMCCSDt and 
IP-CC(t;3)$_{\mathrm{X}}$ calculations 
consisted of the highest-energy $\pi$ orbitals occupied in (SH)$^{-}$.
}
\end{tablenotes}
\end{threeparttable}
\end{table*}
%
\begin{table*}[h]
\centering
\footnotesize
\begin{threeparttable}
\caption{
The total electronic energies obtained using the IP-EOMCCSD(2$h$-1$p$), IP-CR-EOMCC(2,3),
IP-EOMCCSDt, IP-CC(t;3), and IP-EOMCCSD(3$h$-2$p$) approaches for the ground
and low-lying excited states of the $\rm{N_3}$ and NCO molecules,
as described using the DZP[4s2p1d] basis set of Refs.\ \cite{dzp1,dzp2}.
The IP-EOMCCSD(3$h$-2$p$) results are given in hartree and the IP-EOMCCSD(2$h$-1$p$),
IP-CR-EOMCC(2,3), IP-EOMCCSDt, and IP-CC(t;3) data are reported as errors relative
to IP-EOMCCSD(3$h$-2$p$) in millihartree.
The RHF orbitals of the closed-shell $\rm{(N_3)}^{-}$ and $\rm{(NCO)}^{-}$ anion
references were used in the IP-EOMCC calculations for $\rm{N_3}$ and NCO,
respectively. The lowest-energy orbitals that correlate with the 
1s shell of carbon, nitrogen, and oxygen were kept frozen.
}
\label{table5}
\begin{tabular}{llccccccc}
\hline\hline
& & & \multicolumn{2}{c}{IP-CR-EOMCC(2,3)$_{\rm X}$} & & \multicolumn{2}{c}{IP-CC(t;3)$_{\rm X}$\tnote{b}} & \\
\cline{4-5} \cline{7-8}
Molecule\tnote{a} & State & IP-EOMCCSD(2$h$-1$p$) & $\rm{X} = \rm{A}$ & $\rm{X} = \rm{D}$ & IP-EOMCCSDt\tnote{b} & $\rm{X} = \rm{A}$ & \rm{X} = \rm{D} & IP-EOMCCSD($3h\mbox{-}2p$) \\
\hline
N$_{3}$ & ${X\:^{2}}\Pi_{\rm g}$        & 13.078 & 1.509 & $-$0.808 & 1.400 & 0.169 & 0.004 & $-$163.719 333 \\
	   & ${B\:^{2}}\Sigma_{\rm u}^{+}$ & 14.623 & 7.000 & 5.208    & 6.215 & 1.375 & 0.229 & $-$163.560 374
\vspace{1em} \\
NCO & ${X\:^{2}}\Pi$        & 9.587  & 2.374 & 1.273 & 2.205  & 0.205 & 0.062   & $-$167.591 596 \\
    & ${A\:^{2}}\Sigma^{+}$ & 10.921 & 4.994 & 3.236 & 10.089 & 1.690 & $-$0.189 & $-$167.486 358 \\
    & ${B\:^{2}}\Pi$        & 20.154 & 7.044 & 4.504 & 1.926  & 1.123 & 0.863    & $-$167.447 865 \\
\hline\hline
\end{tabular}
\begin{tablenotes}
\footnotesize
\item[a]{
Both the N$_{3}$ and NCO molecules employed linear nuclear geometries, with
N--N, C--N, and C--O bond lengths taken from Ref.\ \cite{ehara},
in which they were optimized for each electronic state using 
the IP SAC-CI SDT-$R$ method.
}
\item[b]{
The active spaces used in the IP-EOMCCSDt and IP-CC(t;3)$_{\mathrm{X}}$ calculations
for $\rm{N_3}$ and NCO consisted of the highest-energy 
$\pi$ orbitals occupied in $\rm{(N_3)}^{-}$ and 
$\rm{(NCO)}^{-}$, respectively.
}
\end{tablenotes}
\end{threeparttable}
\end{table*}
\end{document}